\documentclass[aps,twocolumn]{revtex4}
\usepackage[normalem]{ulem}
\usepackage{bm}
\usepackage[usenames, dvipsnames]{color}
\usepackage{graphicx,amsmath}
\usepackage[colorlinks]{hyperref}
\hypersetup{colorlinks,citecolor=red,linkcolor=blue,urlcolor=blue}

\begin{document}

\title{L\'evy flights confinement in a parabolic potential and fractional quantum oscillator}

\author{E. V. Kirichenko}
\affiliation{Institute of Mathematics and Informatics, Opole
University,\\Oleska 48, 45-052, Opole, Poland}

\author{V. A. Stephanovich}\affiliation{Institute of Physics, Opole
University,\\Oleska 48, 45-052, Opole, Poland}

\date{\today }

\begin{abstract}
We study L\'evy flights confined in a parabolic potential. This has to do with a fractional 
generalization of ordinary quantum-mechanical oscillator problem. To solve the spectral problem
for the fractional quantum oscillator, we pass to the momentum space, where we apply the
variational method. This permits to obtain approximate analytical expressions for eigenvalues and 
eigenfunctions with very good accuracy. Latter fact has been checked by numerical solution of the problem.
We point to the realistic physical systems ranging from multiferroics and oxide heterostructures to quantum chaotic excitons, where obtained results can be used.
\end{abstract}
\maketitle

\section{Introduction}
An introduction of fractional derivatives to describe the non-Gaussian 
phenomena has become a truism. The common knowledge about fractional 
derivatives is that they generate the heavy-tailed, non-Gaussian probability 
densities, both in spatial and temporal patterns. The most prominent example here
is so-called anomalous diffusion \cite{mk2000, mk2004}, intimately related to L\'evy flights 
\cite{mk2000,mk2004,levy,lf1, hud,st}. L\'evy flights constitute a Markovian random process whose probability density function (pdf) is a L\'evy stable law, $f(x,t)$ of index $0<\mu \leq 2$. In the infinite interval $x \in {\mathcal R}$ it is convenient to define the pdf in terms of its characteristic function $f (k,t)$. For the free (i.e. without external potential, so-called untamed) L\'evy flights, the pdf $f(x,t)$ is determined by the fractional Fokker-Planck (FP) equation, see, e.g., Ref. \cite{mk2000}. In dimensionless units (diffusion coefficient and particle mass are sent to unity) it reads
\begin{equation}\label{qo1}
\frac{\partial}{\partial t} f(x,t)=\frac{\partial^\mu}{\partial |x|^\mu}f(x,t)\equiv -|\Delta|^{\mu/2}f(x,t),
\end{equation}
where $|\Delta|^{\mu/2}$ is a one-dimensional (1D) fractional Laplacian, which at $\mu=2$ yields the ordinary one \cite{pod,sam}. We note here that although at $\mu=2$ fractional Laplacian gives ordinary one, the case $\mu=1$ does {\em{not}} correspond to the ordinary first derivative $d/dx$, but rather to -$d/d|x|$ (Riesz fractional derivative \cite{sam,pod}) which is again fractional operator. The definition of the fractional Laplacian reads:
\begin{eqnarray}
-|\Delta|^{\mu/2}f(x)=A_\mu \int_{-\infty}^\infty \frac{f(y)-f(x)}{|y-x|^{1+\mu}}, \label{qo2} \\
A_\mu=\frac{1}{\pi}\Gamma(1+\mu)\sin \frac{\pi \mu}{2}. \label{qo3}
\end{eqnarray}
It is seen that operator \eqref{qo2} is spatially nonlocal with a slowly decaying power-law kernel typical for memory effects in complex systems. One more interesting application of L\'evy processes is so-called fractional quantum mechanics \cite{lask2000, laskbook}, dealing in short with the substitution of ordinary Laplacian with fractional one \eqref{qo2} in the stationary Schr\"odinger equation. The solution of such problem, if exists,  represents the spectrum of a corresponding fractional Hamiltonian. The information about the latter spectrum is very useful as it permits to look for solutions of the fractional FP equations in external potential (i.e. non-free version of the equation \eqref{qo1}) as an expansion over the complete set of orthogonal eigenfunctions of a properly tailored fractional Hamiltonian. Under proper tailoring here we understand the choice of a potential, which is related to that in the corresponding fractional FP equation, which can be regarded as fractional generalization of a Sturm-Liouville operator, see below. It is well-known (see, e.g., book \cite{land3}) that the right choice of the orthogonal base increases dramatically the  corresponding series convergence. In that sense, the spectra of above fractional Hamiltonians, if found (even approximately) analytically, represent the useful orthonormal bases, which can be further employed for the solutions of the problems, dealing with the L\'evy flights confinement in corresponding potentials. 

 In the present paper we focus on the L\'evy flights of arbitrary index $0<\mu \leq 2$ confined in the parabolic potential well, which corresponds to the problem of a fractional quantum harmonic oscillator. To be specific, here we consider the following spectral problem 
\begin{equation}\label{qo4}
-|\Delta|^{\mu/2}\psi_{i\mu}(x) + \frac{x^2}{2}\psi_{i\mu}(x)=E_{i\mu}\psi_{i\mu}(x).
\end{equation}
Here we adopt the units $\hbar =m=\omega=1$, where $m$ and $\omega$ are the mass and frequency of oscillating particle respectively. Also, $\psi_{i\mu}(x)$ is the $i$-th eigenfunction of a fractional quantum harmonic oscillator having the eigenenergy $E_{i\mu}$ for any specific $\mu$ value. 

There are several examples of the problems resembling \eqref{qo4} but not exactly similar to it. First one has been considered by Laskin \cite{laskbook}. This problem is little more general than \eqref{qo4} as it has the potential energy $|x|^\beta$, $1 <\beta \leq 2$, related to the quark confinement theory.  Although the problem has been formulated both in coordinate and momentum spaces, its solution has been represented in semiclassical case only. Below we will show that our methods can be well applied to solve this problem in "purely quantum" (i.e. not semiclassical) case.  

Other example has been considered in the papers \cite{mf99,mbk99}, where the fractional FP equation for L\'evy Ornstein-Uhlenbeck process has been solved by a separation ansats. This ansats decomposes the initial equation on time and coordinate-dependent parts. Latter part admits the complete solution in terms of the spectrum of a fractional Fokker-Planck operator \cite{mf99}. General form (for arbitrary potential function $U(x)$) of such operator reads in the above dimensionless variables 
\begin{equation}\label{fpp1}
{\hat L} _{FP}\psi=\frac{\partial}{\partial x}\left(\psi\frac{dU}{dx}\right)+|\Delta|^{\mu/2}\psi.
\end{equation}
For $U(x)=x^2/2$, the first term in Eq. \eqref{fpp1} has the form $x\psi'(x)+\psi(x)$, which is obviously different form the second term of the Schr\"odinger operator \eqref{qo4}. The operator \eqref{fpp1} can be regarded as fractional generalization of Sturm-Liouville operator. Also, seminal Landau-Teller model of molecular collisions \cite{land} and its fractional generalization \cite{carati} deals with classical oscillator equation related to corresponding sound waves. It is tempting to quantize this problem in terms of fractional Schr\"odinger equation.

Here we are going to solve the spectral problem for the fractional quantum harmonic oscillator of arbitrary $\mu$. As this problem resides on the whole real axis, the most profitable way to solve it is to use momentum space, where the problem becomes local. Namely, in momentum space we are dealing with ordinary (i.e. with second spatial derivative, stemming from the potential in $x$ space) Schr\"odinger equation, which permits to apply all known approaches (like variational and well-developed numerical) to solve the spectral problem. To find the eigenfunctions in the $x$ space, we perform 
inverse Fourier transformation. Specifically, here we shall utilize the variational approach to find the spectral solution of the problem \eqref{qo4}. It is well-known (see, e.g. Ref. \cite{bershub}) that variational methods work for self-adjoint operators, which is the case for ordinary (i.e. without fractional derivatives) quantum mechanics. Below we shall see, that operator \eqref{qo4} in $k$ space is self-adjoint. This follows from the theorem 1.1 on p. 50 of the Ref. \cite{bershub}, which will be discussed quantitatively below. We postpone the studies of self-adjointness of the other (than \eqref{qo4}) fractional quantum mechanics Hamiltonians (along with their variational treatment) to the future publications. 

\section{General formalism}

Our aim is to solve the spectral problem \eqref{qo4} for fractional quantum harmonic oscillator. In the article \cite{yama}, we adopted the method for solution of the spectral problems like \eqref{qo4}. The idea is to expand the solution in the complete orthonormal set of the eigenfunctions, formed by the solution of corresponding "ordinary" (i.e. that for $\mu=2$) quantum-mechanical problem. In our case it is an oscillator, whose wave functions (in our units) are given by the well-known expressions, see, e.g. Ref. \cite{land3}
\begin{equation}\label{vfmu2}
\psi_{n,\mu=2}(x)=\frac{H_n(x)e^{-x^2/2}}{\pi^{1/4}\sqrt{2^n\cdot n!}},
\end{equation}
where $H_n(x)$ are Hermite polynomials of $n$-th order \cite{abr}. It turns out, however, that the matrix method, adopted in Ref. \cite{yama} for our problem, converges extremely slowly so that large (around $10^4\times 10^4$) matrices are to be diagonalized. This, along with quite long time, needed to calculate each matrix element, renders this method unsuitable for our present problem. Rather, here we utilize the Fourier techniques, considering the problem in momentum space. The advantage is that in $k$ - space the problem turns into ordinary Schr\"odinger equation with a large arsenal of tools for its solution.
 
In the momentum space, the equation \eqref{qo4} assumes especially simple form
\begin{eqnarray}
\mathcal{H}_k\psi_{i\mu}(k)=E_{i\mu}\psi_{i\mu}(k),\label{qo5a} \\
\mathcal{H}_k= -\frac{1}{2}\frac{d^2}{dk^2}+\frac{1}{2}|k|^\mu. \label{qo5b}
\end{eqnarray}

\begin{figure}[t]
\begin{center}
\includegraphics*[width=1.1\columnwidth]{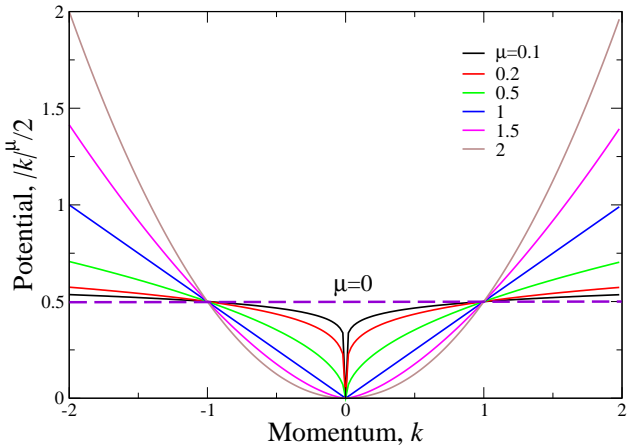}
\end{center}
\caption{Potential in the equation \eqref{qo4} for different $\mu$, shown in the legend. The potential for $\mu=0$ is shown by dashed horisontal line. }
\label{fig:pot}
\end{figure}

The equation \eqref{qo5a} represents Schr\"odinger equation with potential $|k|^\mu/2$. The plots of latter potential at different $\mu$  are shown in Fig. \ref{fig:pot}. It is seen that at $\mu=0.1$ and 0.2 the potential differs from that for $\mu=0$ (dashed horisontal line) only near $k=0$, where it has nonanalytical behavior with infinite derivative. The same behavior persists up to $\mu=1$, at which point the derivative at $k=0$ is constant as we have straight line in this case. At $1 \leq \mu \leq 2$ we have parabola-like curves with zero derivative at $k=0$. At $\mu=2$ we recover the parabolic potential for ordinary quantum oscillator in $k$ space. 
The main feature of the potential \eqref{qo5b} is that except for $\mu=0$ (which is not included in the fractional Laplacian domain) it grows to infinity at $k \to \pm \infty$. This means that we should have the discrete spectrum for the fractional quantum oscillator at the entire domain $0<\mu \leq 2$. Latter fact implies, in turn, that the wave functions should be localized in $k$ - space.

To analyse the character of wave functions localization in $k$ - (and eventually in the $x$) space, we should find the large $k$ asymptotics of the  $\psi(k)$. For that we observe that at large $k$ we can neglect the term $E\psi$ in the right-hand side of Eq. \eqref{qo5a}. This generates the equation for large $k$ asymptotics in the form $\psi''(k)=|k|^\mu \psi$. The spatially decaying solution to this equation can be found (Ref. \cite{pol}, see also \cite{my11}) to be proportional to $k^{1/2}K_\nu(u)$, where $\nu=1/(\mu+2)$ and $u=2\sqrt{2}|k|^{1+\mu/2}/(\mu+2)$. Here $K_\nu(x)$ is modified Bessel function with the following large $x$ asymptotics $K_\nu(x \to \infty)\approx (\pi/(2x))^{1/2}e^{-x}$ \cite{abr}. Substitution of latter asymptotics into the expression $\psi \sim  k^{1/2}K_\nu(u)$ yields
\begin{equation}\label{qo6}
\psi_{i\mu}(k \to \infty) \sim |k|^{-\mu/4}\exp\left[-\frac{2\sqrt{2}}{\mu+2}|k|^{1+\mu/2}\right].
\end{equation}
It can be shown, that the main (i.e. the largest at $k \to \infty$) term of the second derivative 
$\psi''(k \to \infty)$ is really proportional to $|k|^\mu \psi$. In the case of $\mu=2$ (ordinary quantum oscillator) this reproduces well known result $\psi''(\mu=2,k \to \infty)=k^2\psi$ \cite{land3}. It is seen that for all $0<\mu \leq 2$ the wave function \eqref{qo6} is well localized: even at $\mu=0$ $\psi(k \to \infty) \sim e^{-|k|\sqrt{2}}$, i.e. decays exponentially. At $\mu>0$ the decay is faster and at $\mu=2$ we arrive at correct asymptotics $e^{-k^2/\sqrt{2}}$ corresponding to ordinary quantum oscillator. Good localization of the $\psi$ functions in $k$ space yields their absolute integrability for all $0<\mu \leq 2$ (i.e. the integral $\int_{-\infty}^\infty |\psi(k)|dk$ is finite) and hence (by Riemann-Lebesgue lemma, see e.g. \cite{intr}) the localization of the wave functions $\psi(x)$ in coordinate space.

\section{Variational treatment}

The equation \eqref{qo5a} can be solved analytically in two cases. First corresponds to $\mu=2$ and comprises ordinary quantum oscillator \cite{land3}. The second one corresponds to $\mu=1$ and admits the exact solution in terms of Airy functions, see Ref. \cite{my11} and references therein. The solution for the rest of L\'evy indices can be found, generally speaking, only numerically. Here we suggest the approximate analytical method to find the spectrum of the operator \eqref{qo5b} for all $0<\mu\leq 2$. This method is based on variational solution of the Schr\"odinger equation \eqref{qo5a}.
To this end, we should establish the self-adjointness of the operator \eqref{qo5b}. This can be done on the base of the theorem 1.1 of Ref. \cite{bershub} (see p.50 of this book), which states, that for operator \eqref{qo5b} to be self-adjoint, it is necessary and sufficient that the potential  $v(x) \geq -Q(x)$ such that $\int_{-\infty}^{\infty}[Q(2x)]^{-1/2}dx=\infty$. The function $Q(x)$ should be positive even continuous non-decreasing function on the whole real axis. The Figure \ref{fig:pot} shows that the simplest choice of such function is any positive constant, $Q(x)>0=const$. Such choice guarantees the fulfillment of the above theorem conditions and proves that Hamiltonian \eqref{qo5b} is essentially self-adjoint. This means, in turn, that the variational principle of quantum mechanics \cite{land3} can be well applied for the approximate solution of the equation \eqref{qo5a}. The more general quantum oscillator problem, considered by Laskin \cite{laskbook}, contains fractional derivative of index $\beta$ (see above) also in momentum space. To prove the self-adjointness of the corresponding Hamiltonian, we should follow the proof of the theorem 1.1 from Ref. \cite{bershub}. This proof is based essentially on the analysis of wave function asymptotics at $x \to \pm \infty$. Our analysis shows that at $1<\beta \leq 2$ the wave function decays sufficiently fast so that the corresponding Hamiltonian operator is self-adjoint. In this case, the obtained asymptotics of the wave function should be used in trial wave functions for variational treatment. 

The asymptotics \eqref{qo6} can be employed to construct the trial wave functions $\psi_{i\mu}(k)$ for any $\mu$ from the domain $0<\mu\leq 2$. As usual, the variational solution of the spectral problem \eqref{qo5a} should minimize the energies
\begin{equation}\label{av1}
 W_{i\mu}=\int_{-\infty}^\infty \psi_{i\mu}^*(k)\ {\mathcal{H}_k}\psi_{i\mu}(k)dk.
\end{equation}
Here ${\mathcal{H}_k}$ is the Hamiltonian \eqref{qo5b} and $W_{i\mu}$ are the variational approximations of the eigenenergies $E_{i\mu}$. Normally $W_{i\mu} \geq E_{i\mu}$. As wave functions $\psi_{i\mu}(k)$ can be chosen to be real, the complex conjugation sign in Eq. \eqref{av1} is not necessary. Also, since functions $\psi_{i\mu}(k)$ are well localized (see asymptotics \eqref{qo6}), the expression \eqref{av1} could be rendered (by the integration by parts in the first term of \eqref{qo5b}) to the more convenient form
\begin{equation}\label{av2}
W_{i\mu}=\int_{-\infty}^\infty \bigg[(\psi'_{i\mu})^2+|k|^\mu \psi_{i\mu}^2\bigg]dk,
\end{equation}
where $\psi'=d\psi/dk$. 

We look for the trial functions on the base of asymptotics \eqref{qo6} and oscillational theorem (see, e.g.,\cite{land3}), stating that the wave function of $i$-th state has $i$ nodes. In other words, the ground state wave function $\psi_0$ has no nodes, $\psi_1$ has one node ets. This implies, in turn, the mutual orthogonality of the trial functions $\psi_{i\mu}$. In the simplest possible form the trial functions read
\begin{eqnarray}
\psi_{0\mu}=A_{0\mu}e^{-a_{0\mu}|k|^{1+\mu/2}}, \ \psi_{1\mu}=A_{1\mu}ke^{-a_{1\mu}|k|^{1+\mu/2}},\nonumber \\
\psi_{2\mu}=A_{2\mu}(b_{0\mu}+b_{2\mu}k^2)e^{-a_{2\mu}|k|^{1+\mu/2}},...\label{av3}
\end{eqnarray}
Here $a_{i\mu}$ $(i=0,1,2)$ and $b_{i\mu}$ ($i=0,2$) are variational parameters, while $A_{i\mu}$ are normalization coefficients, found from the obvious condition 
\begin{equation}\label{orm}
\int_{-\infty}^\infty \psi_{i\mu}^2(k)dk=1.
\end{equation} 
The condition \eqref{orm} relates $A_{i\mu}$ to $a_{i\mu}$ and $b_{i\mu}$. With respect to normalization condition \eqref{orm} we now find variational parameter $a_{0\mu}$ from the minimum of the functional \eqref{av2}. The parameter $a_{1\mu}$ should be found from the minimum of $\eqref{av2}$ with additional condition of orthogonality $\int_{-\infty}^\infty\psi_{1\mu}\psi_{0\mu}dk=0$. Latter condition is satisfied automatically as can be seen from Ex. \eqref{av3}. The parameters $b_{0,1\mu}$ are related to $a_{2\mu}$ by two orthogonality conditions  $\int_{-\infty}^\infty\psi_{2\mu}\psi_{0\mu}dk=0$ and  $\int_{-\infty}^\infty\psi_{2\mu}\psi_{1\mu}dk=0$. Then, $a_{2\mu}$ is found from the minimum of the energy functional \eqref{av2}. Such procedure can be done for all wave functions of higher excited states $i>2$, giving the approximate spectrum of the operator \eqref{qo5b}. Note that the substitution of found 
$a_{0\mu}$ into $W_{0\mu}$ \eqref{av2} gives the approximate value of the ground state energy for all $\mu$, the same with $a_{1\mu}$ gives the energy of first excited state $W_{1\mu}$ and so on for higher eigenenergies. This shows the advantage of variational method, which permits to obtain the approximate analytical expressions for the eigenvalues and eigenfunctions of the operator \eqref{qo5b} for all $0<\mu\leq 2$. Below we check the accuracy of our variational method by comparison of its results with numerical solution of the spectral problem \eqref{qo5a}. In order to improve the accuracy, we should increase the number of variational parameters.

Substitution of the trial function $\psi_{0\mu}$ into the integral \eqref{av2} with subsequent minimization over $a_{0\mu}$ yields
\begin{equation}\label{a0}
(a_{0\mu})_{min}=\frac{\sqrt{2\mu}}{2+\mu}.
\end{equation}
Further substitution of this value to the result for $W_{0\mu}$ generates the approximate value of the ground state energy for arbitrary $\mu$
\begin{equation}\label{w0}
(W_{0\mu})_{min} \approx E_{0\mu}=\frac{\mu}{4}\ \frac{\Gamma\left(\frac{\mu}{2+\mu}\right)}{\Gamma\left(\frac{2}{2+\mu}\right)}\left(\frac{2\sqrt{2\mu}}{2+\mu}\right)^{-\frac{2\mu}{2+\mu}}.
\end{equation}
It is seen that $(W_{0\mu})_{min}$ gives correct value 1/2 of the ground state energy for $\mu=2$, corresponding to the ordinary quantum oscillator with the spectrum $E_n=n+1/2$ in our units. Below we shall see, that for the case $\mu=0$ all the spectrum shrinks into a single value $E_0=1/2$, which is also obtained correctly from the expression \eqref{w0}. The entire $\mu$ dependence \eqref{w0} will be plotted below and compared with numerical solution.

\begin{figure}[t]
\begin{center}
\includegraphics*[width=1.1\columnwidth]{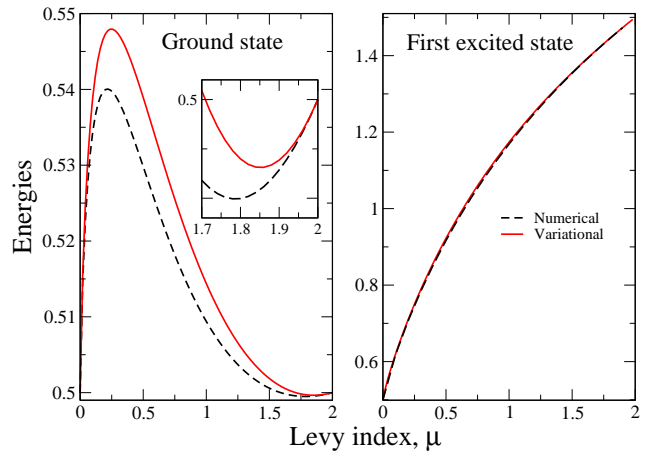}
\end{center}
\caption{Comparison of variational (Eqs \eqref{w0} and \eqref{w1}, solid lines) and numerical values (dashed lines) of the ground and first excited states energies as functions of L\'evy index $\mu$. Inset to left panel details the behavior of the curves at $\mu \to 2$. Mind the different vertical scales in left (ground state) and right (first excited state) panels.}
\label{fig:sr}
\end{figure}

The same procedure with $\psi_{1\mu}$ gives that $(a_{1\mu})_{min}=(a_{0\mu})_{min}$, which is given by Ex. \eqref{a0}. The variational expression for the energy of the first excited state reads
\begin{eqnarray}
(W_{1\mu})_{min} \approx E_{1\mu}=\frac{(2+\mu)(4+\mu)}{\mu}\times \nonumber \\
 \frac{\Gamma\left(\frac{2}{2+\mu}\right)}{\Gamma\left(\frac{6}{2+\mu}\right)}\left(\frac{2\mu}{(2+\mu)^2}\right)^{\frac{2}{2+\mu}}\ 2^{-\frac{4(1+\mu)}{2+\mu}}. \label{w1}
\end{eqnarray}
We see that at $\mu=2$ the expression \eqref{w1} gives the correct answer 3/2. At the same time, at $\mu \to 0$ we have removable divergence $\lim_{\mu \to 0}\frac{\mu^{2/(2+\mu)}}{\mu}=1$, while the rest of the expression \eqref{w1} gives $E_{1,\mu=0}=\Gamma(1)/\Gamma(3)=1/2$, i.e. once more the correct answer. The dependence \eqref{w1} will also be plotted below and compared with numerical results. We will see that the expressions \eqref{w0} and \eqref{w1} give very good approximate expressions for ground and first excited state energies of a fractional quantum oscillator for the entire $\mu$ domain. Within the suggested variational approach, any desired energy level $E_{i\mu}$ can be evaluated analytically, although the calculations for higher excited states become very cumbersome.

\section{Numerical analysis}

We begin with the analysis of the system spectrum for arbitrary $\mu$. Table \ref{yt} shows five lowest eigenenergies of the operator \eqref{qo5b}, calculated numerically with the help of Mathematica routine \texttt{NDEigensystem}. It is seen that at $\mu=2$ we have the spectrum $E_{i,\mu=2}=i+1/2$ of the ordinary quantum oscillator, while at $\mu$ decrease the spectrum deviates from $i+1/2$ and at $\mu \to 0$ (for instance at $\mu=0.1$) all spectrum is concentrated around single value $E_0=1/2$. Note, that the same regularities take place also for fractional quantum well \cite{yama}. 
\begin{table*}[t]
\begin{tabular}{ccccccc}
\hline \hline
Number of state, $i$ & 0 & 1 & 2 & 3 & 4 & 5 \\ \hline 
$\mu=0.1$  &0.534418 & 0.617864 & 0.643554& 0.667326& 0.681551& 0.696129 \\ 
$\mu=0.5$  &0.529809 & 0.916697& 1.10501& 1.27532& 1.40276& 1.52559 \\ 
$\mu=1.0$  &0.509396 & 1.16905 & 1.6241 & 2.04398 & 2.41005 & 2.76028 \\ 
$\mu=1.5$  &0.500592 & 1.35405 & 2.08857 & 2.79283 & 3.46141 & 4.11343 \\
$\mu=1.8$  &0.499498&1.44520 & 2.34152 & 3.22291 & 4.08849 &  4.94536\\ 
$\mu=2.0$  &0.5 &1.5 & 2.5 & 3.5 & 4.5 & 5.5\\ \hline \hline
\end{tabular}
\caption{Five lowest eigenstates of the fractional quantum harmonic oscillator for different $\mu$, obtained numerically.}\label{yt}
\end{table*} 
As we can reproduce numerically the spectrum of the operator \eqref{qo5b} for arbitrary $\mu$, we are now in a position to compare the variational expressions \eqref{w0} and \eqref{w1} with corresponding numerical values. Such comparison is reported in Fig. \ref{fig:sr}. As it is the case for variational method, in both panels the variational curves lie higher than numerical ones as variational energy should be larger than its exact (in our case numerical) value \cite{land3}.

\begin{figure}[t]
\begin{center}
\includegraphics[width=1.1\columnwidth]{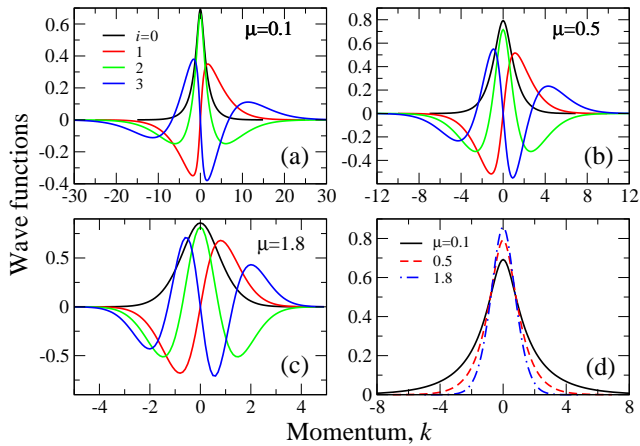}
\end{center}
\caption{Panels (a) - (c): four first wave functions of the fractional quantum oscillator in $k$ - space for different $\mu$, shown in the legends. Number of state $i$ is also shown. Panel (d) compares the ground state ($i=0$) functions for different  $\mu$ (legend). The different horisontal scales in the panels (a) - (c) reflect the character of decay at different $\mu$.}
\label{fig:vf}
\end{figure}

\begin{figure}[t]
\begin{center}
\includegraphics[width=1.1\columnwidth]{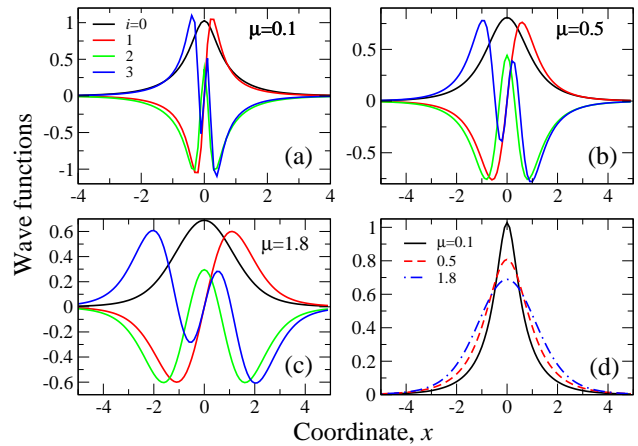}
\end{center}
\caption{Same, as in Fig. \ref{fig:vf}, but in coordinate space. In the panel (d), now  fastest decaying function is that for 
$\mu=0.1$.}
\label{fig:vfx}
\end{figure}
 
It is seen that the agreement is much better for the first excited state, where the variational and numerical curves are indistinguishable in the scale of the plots. The average error in this case is only 0.3\%. Our analysis shows that the same good accuracy occurs for higher excited states also. This means that the variational expressions like \eqref{w1} can be regarded as "almost exact" for the states with $i \geq 1$. This is because the energies of excited states are much higher then ground state one so that in the scale of right panel of Fig. \ref{fig:sr}, the ground state curves would lie in the narrow strip near $x$ axis. The second factor, influencing such good variational approximation with the only one parameter is that the excited state energies are monotonous functions of $\mu$. On the contrary, the ground state energy is non-monotonous function of $\mu$ as it begins (at $\mu=0$) and ends (at $\mu=2$) at the same value 0.5. It is seen from left panel of the Fig.  \ref{fig:sr}, that the largest error about 1.5\%  occurs near the curve maximum, i.e. around $\mu_{max}\approx 0.2476$ for variational curve. Thus we can safely assert that with the accuracy not higher than 1.5\% the analytical expression \eqref{w0} approximates the numerical curve. This means that the approximation for ground state energy is also not bad at all for the trial function with only one parameter.  To improve the accuracy, the consideration of trial functions with more adjustable parameters is necessary.

One more interesting (although tiny) feature of $E_{0\mu}$ curves is shown in the inset to the left panel of Fig. \ref{fig:sr}. Namely, both numerical and variational curves have minima at $\mu=1.8 - 1.85$ and then approach the asymptotic value  $E_{0\mu}=0.5$ at $\mu=2$. This minute difference between the ground state energies of the ordinary quantum oscillator ($\mu=2$) and its "almost ordinary" counterpart ($\mu \approx 1.85$) means that the fractional quantum oscillator with L\'evy index about 1.85 has lower ground state energy than ordinary one. Our numerical calculations of the spectrum for $\mu=1.85$ show that the energies for this case lie lower than those for $\mu=2$. We have for the first four eigenvalues: $E_0=0.499543$ (against 0.5 at $\mu=2$), $E_1=1.4593$ (1.5 for $\mu=2$), $E_2= 2.38189$ (2.5 for $\mu=2$), $E_3=3.29293$ (3.5 for $\mu=2$).

The numerical wave functions in $k$ - space are reported in Fig. \ref{fig:vf} for different values of $\mu$. It is seen that the oscillation theorem holds, i.e. the wave function of the state number $i$ has exactly $i$ nodes. It can be checked that functions are normalized, i.e. they obey condition \eqref{orm}. It is seen from panels (a) - (c) that functions for different $\mu$ has different decay rates. These rates are dictated by the asymptotics \eqref{qo6}. The decay rates are smaller for $\mu \to 0$ (at $\mu=0$ all wave functions do not decay at all) and larger for $\mu \to 2$, tending to those of conventional quantum oscillator. Note that variational wave functions \eqref{av3} (with respect to minimizing parameters defined by the expression \eqref{a0}) do not differ from those in Fig.  \ref{fig:vf} (a) - (c) in the scales of the plots. This means that variational expressions for wave functions also give very good approximations for exact (i.e. numerical) ones. 

The wave function in the coordinate space can be obtained by the inverse Fourier transform $\psi(x)=\frac{1}{\sqrt{2\pi}}\int_{-\infty}^\infty \psi(k)e^{ikx}dk$. It is reported in Fig. \ref{fig:vfx}. Similar to the case of momentum space, the oscillation theorem also holds and the wave functions are normalized to unity. The only difference between $k$ - and $x$ spaces is that the fastest decaying function is now that for $\mu=0.1$. This is because for $\mu=0$ the all wave functions merge into single Dirac $\delta$ function. Latter also follows from the fact that in $k$ - space for $\mu=0$ all functions merge into a constant. Accordingly, the slowest decaying function is that for $\mu=1.8$. At $\mu=2$ we once more have the ordinary quantum oscillator wave functions, given by the expression \eqref{vfmu2}.

\section{Outlook}

In summary, we have studied the spectral problem for a fractional quantum harmonic oscillator with
arbitrary L\'evy index $0<\mu\leq 2$. As this problem resides in the whole real axis, the most profitable 
way of solution is to pass to momentum space. In the latter space, due to harmonicity of the potential, the problem reduces to the ordinary (i.e. that with second spatial derivative in 1D case) quantum-mechanical one with the potential
$|k|^\mu/2$ containing L\'evy index $\mu$. Having the proof \cite{bershub} of the self-adjointness of the Hamiltonian \eqref{qo5b}, we can safely apply the variational method of quantum mechanics for the problem under consideration. For the anharmonic (in coordinate space) potentials it is not clear if the variational method works in the case of fractional Laplacian. The same is relevant to the other problems of fractional quantum mechanics \cite{laskbook} and for quantum oscialltor, considered there, in particular. The work in this direction should be based on the proof of self-adjointness of the corresponding Hamiltonian operator. This proof, in turn, is based on the analysis of wave function decay character at infinities. The consideration of this interesting class of problems is currently underway. If the variational treatment works for arbitrary potentials in the fractional Scr\"odinger equation, many problems can get their approximate (as we see above, the approximation is generally very good) analytical solutions for arbitrary $\mu$. 

As we have mentioned above, the solution of spectral problems for fractional Hamiltonians can be regarded as creation of the orthonormal bases, in which the solutions of fractional FP equations could be expanded. This method can be considered as more general since only few potentials $U(x)$ in the fractional FP equations admit the exact solution. One of the cases had been considered in Ref. \cite{mf99}, where $U(x)=x^2/2$, see also above. The solution had been done in momentum space, where the explicit form of the spectral problem reads in our dimensionless units
\begin{equation}\label{nuka}
k\frac{d}{dk}\psi_n(k)+|k|^\mu\psi_n(k)=\lambda_n\psi_n(k),
\end{equation}
where 
\begin{equation} \label{kuka}
\psi_n(k)=c_n|k|^{\mu n}\exp\left[-\frac{|k|^\mu}{\mu}\right]
\end{equation}
is the eigenfunction (with $c_n$ being normalization constant), corresponding to $n$-th eigenvalue $\lambda_n=\mu n$. It is seen that the "kinetic part" (i.e. that containing derivatives) of the operator \eqref{nuka} is different from that in our expression \eqref{qo5b}. This, actually, is the reason, why the function \eqref{kuka} (and similar {\em{ans\"atze}}) does not satisfy our equation \eqref{qo5a}. Our analysis shows that the equation \eqref{nuka} can be solved by the expansion over the orthonormal set \eqref{av3}, however, the convergence of corresponding series will be worse then that in Eq. (29) of Ref. \cite{mf99}, realizing the expansion over set \eqref{kuka}. This is because the set \eqref{kuka} represent the rare case of exact solution. On the other hand, the L\'evy flights in nonlinear potentials (see, e.g. \cite{chechkin, my11a}) as a rule cannot be solved exactly, while the expansions over complete sets (either exact or obtained variationally like \eqref{av3}) generated by fractional Hamiltonians, can be regarded as a feasible way of such problems solution.

The developed formalism for fractional quantum harmonic oscillator can be applied to the calculations 
of the properties of real physical systems, where disorder (like lattice imperfections and/or impurities) influences phonon and electron spectra of a substance, leading to non-Gaussian distribution of the internal electric, magnetic and elastic fields. Challenging example here is electric and magnetic properties of multiferroics, where ferroelectric and magnetic orders coexist \cite{multi}.  The non-Gaussian statistics due to disorder and frustration plays an important role in these substances \cite{st1,st2,st3} and we are applying now our formalism to explain unusual experimental data in them. In this context it would be also interesting to consider the fractional generalization of the problem of spatial quantum oscillator  (i.e. particle with potential energy $U(r)=r^2/2$ in our units; $r^2=x^2+y^2+z^2$) \cite{land3}, which arises naturally in above substances as well as in other realistic 2D and 3D physical systems. In dealing with these systems, we should use  multi-dimensional generalization of the fractional Laplacian \eqref{qo3} (see Ref. \cite{my13} and references therein)
\begin{eqnarray}
-|\Delta|^{\mu/2}f({\bf x})=A_{\mu, d} \int \frac{f({\bf u})-f({\bf x})}{|{\bf u}-{\bf x}|^{\mu +d}}, \label{iu1} \\
A_{\mu, d}=\frac{2^\mu\Gamma\left(\frac{\mu+d}{2}\right)}{\pi^{d/2}|\Gamma(-\mu/2)|}, \label{iu2}
\end{eqnarray}
where $d$ is space dimensionality. In this case the spectral problem for 3D quantum fractional oscillator reads
\begin{equation} \label{iu3}
-|\Delta|^{\mu/2}\psi_{i\mu}({\bf r})+\frac 12 (x^2+y^2+z^2) \psi_{i\mu}({\bf r})=E_{i\mu}\psi_{i\mu}({\bf r}),
\end{equation}
where ${\bf r}$ is now 3D vector and the other notations are same as those in Eq. \eqref{qo4}. Similar to the considered case of 1D oscillator, this problem resides in the whole space. This means, that once more it is convenient to pass to the  momentum space. With respect to the fact that in momentum space the operator \eqref{iu1} is simply $|{\bf k}|^\mu$ (where ${\bf k}$ is $d$ - dimensional momentum vector), in this space the equation \eqref{iu3} renders to the form
\begin{equation}\label{iu4}
-\Delta_{\bf k} \psi_{i\mu}({\bf k})+k^\mu \psi_{i\mu}({\bf k})=E_{i\mu}\psi_{i\mu}({\bf k}),
\end{equation}
where $k=|{\bf k}|$, $\Delta_{\bf k}\equiv \frac{\partial^2}{\partial k_x^2}+\frac{\partial^2}{\partial k_y^2}+\frac{\partial^2}{\partial k_z^2}$ is the ordinary Laplacian in ${\bf k}$ space. After usual decomposition $\psi_{i\mu}({\bf k})=R_{il}(k)Y_{lm}(\theta, \varphi)$ ($Y_{lm}$ are spherical harmonics and $l,m$ are orbital and magnetic quantum numbers respectively) \cite{land3}, we obtain following fractional Scr\"odinger equation for the radial part $R_{il}(k)$
\begin{equation}\label{iu5}
\frac{d^2R_{il}}{dk^2}+\frac 2k \frac{dR_{il}}{dk}+\left[2E_{i\mu}-\frac{l(l+1)}{k^2}-k^\mu\right]R_{il}=0.
\end{equation}
The equation \eqref{iu5} can also be solved variationally, giving the approximate (but of very good accuracy) spectrum of the 3D fractional oscillator. This spectrum can furthter be used in the calculation of partition function and other thermodynamic characteristics of the above systems. The work on this interesting problem is underway. 

Other object to apply the solutions of fractional quantum mechanical problems is oxide interfaces \cite{ohm,bvb}, where non-Gaussian quantum fluctuations occur both in phonon and electron spectra due to specific potential at the interface \cite{du1,du2,du3}. Here, both the above results on 1D fractional quantum oscillator and those for quantum well \cite{yama} can be well applied. 

Finally we mention one more interesting physical problem regarding the onset of chaos in the excitons (described by the quantum mechanical model of hydrogen atom, see, e.g. \cite{ash}) due to Rashba spin-orbit interaction \cite{br}. This problem turns out to be extremely important for perovskite substances, used in photovoltaics  \cite{pho}, where the above chaos can adversely influence devices functionality. While in the classical case the chaotic electron trajectories in an exciton have been clearly revealed \cite{she1}, in the quantum case only weak effects like level repulsion (but not non-Poissonian energy level statistics, see e.g. \cite{rei}) were seen \cite{she2}. We speculate that the introduction of fractional derivatives in the corresponding 2D Schr\"odinger equation can highlight the quantum chaotic features, which are actually important for photovoltaic devices functionality. This problem can be formulated in the form of equation \eqref{iu3} but with Coulomb potential. The transition to momentum space is also possible, but it is much more laborious then in the case of 3D oscillator \eqref{iu3}. This means that the solution of this problem turns out to be rather involved so that we should opt to the numerical methods.

\begin{acknowledgements}
This work is supported by the National Science Center in Poland as a research project 
No. DEC-2017/27/B/ST3/02881. 
\end{acknowledgements}

\end{document}